\documentclass[draftclsnofoot, 12pt, onecolumn]{IEEEtran}
\usepackage{bbm,amssymb,amsmath,amsthm,amsbsy,mathrsfs}
\usepackage{graphicx,epstopdf,epsfig,psfrag,url,cite,color,multirow,float,makecell}
\usepackage[small,bf]{caption}
\usepackage[font=footnotesize]{subfig}
\usepackage{algorithm,algorithmic}
\usepackage[T1]{fontenc}
\usepackage[latin9]{inputenc}
\usepackage{fixltx2e}
\usepackage{array,multirow,graphicx}
\usepackage{color}
\usepackage{cite}
\usepackage[x11names,dvipsnames,table]{xcolor} 
\usepackage{colortbl}
\usepackage{lipsum}
\definecolor{mygray}{gray}{0.6}
\definecolor{myblue}{rgb}{0.8,0.85,1}
\usepackage{array}
\newcolumntype{L}[1]{>{\raggedright\let\newline\\\arraybackslash\hspace{0pt}}m{#1}}
\newcolumntype{C}[1]{>{\centering\let\newline\\\arraybackslash\hspace{0pt}}m{#1}}
\newcolumntype{R}[1]{>{\raggedleft\let\newline\\\arraybackslash\hspace{0pt}}m{#1}}

\usepackage{array, tabularx, boldline}
\usepackage{eurosym}
\usepackage{amstext} 
\usepackage{cellspace}

\newcommand{\bm}[1]{\boldsymbol{#1}}

\newtheoremstyle{mythm}
{\topsep}   
{\topsep}   
{\itshape}      
{0pt}       
{\bfseries} 
{:}         
{5pt plus 1pt minus 1pt}    
{\thmname{#1}\thmnumber{ #2}\thmnote{ (#3)}}
\theoremstyle{mythm}

\begin{document}

\title{Optimization-driven Machine Learning for Intelligent Reflecting Surfaces Assisted Wireless Networks}

\author{
\IEEEauthorblockN{Shimin Gong, Jiaye Lin, Jinbei Zhang, Dusit Niyato, Dong In Kim, and Mohsen Guizani}
\thanks{Shimin Gong and Jiaye Lin are with the School of Intelligent Systems Engineering, Sun Yat-sen University, China (email: \{gongshm5, linjy98\}@mail.sysu.edu.cn). Jinbei Zhang is with the School of Electronics and Communication Engineering, Sun Yat-sen University, China (email: zhjinbei@mail.sysu.edu.cn). Dusit Niyato is with the School of Computer Science and Engineering, Nanyang Technological University, Singapore (email: dniyato@ntu.edu.sg). Dong In Kim is with the Department of Electrical and Computer Engineering, Sungkyunkwan University (SKKU), Suwon, South Korea (e-mail: dikim@skku.ac.kr). Mohsen Guizani is with the Department of Electrical and Computer Engineering, University of Idaho, Moscow, USA (email: mguizani@ieee.org).
}
}

\maketitle
\thispagestyle{empty}

\begin{abstract}
Intelligent reflecting surface (IRS) has been recently employed to reshape the wireless channels by controlling individual scattering elements' phase shifts, namely, passive beamforming. Due to the large size of scattering elements, the passive beamforming is typically challenged by the high computational complexity and inexact channel information. In this article, we focus on machine learning (ML) approaches for performance maximization in IRS-assisted wireless networks. In general, ML approaches provide enhanced flexibility and robustness against uncertain information and imprecise modeling. Practical challenges still remain mainly due to the demand for a large dataset in offline training and slow convergence in online learning. These observations motivate us to design a novel optimization-driven ML framework for IRS-assisted wireless networks, which takes both advantages of the efficiency in model-based optimization and the robustness in model-free ML approaches. By splitting the decision variables into two parts, one part is obtained by the outer-loop ML approach, while the other part is optimized efficiently by solving an approximate problem. Numerical results verify that the optimization-driven ML approach can improve both the convergence and the reward performance compared to conventional model-free learning approaches.
\end{abstract}
\begin{IEEEkeywords}
Machine learning, intelligent reflecting surface, optimization-driven deep reinforcement learning.
\end{IEEEkeywords}

\section{Introduction}

Recently, the intelligent reflecting surface (IRS) has been introduced in wireless communications to deliberately configure the channel propagation characteristics in favor of information transmission~\cite{irsurvey}. It is composed of a large array of passive scattering elements. Each element can induce individual phase change to the incident RF signals by controlling its operating state. By joint phase control of all scattering elements, namely, passive beamforming, the strength and direction of the reflected signals can be arbitrarily tuned to create desirable channel conditions. As such, the radio environment is turned into a smart space that can be reconfigured and optimized to improve the network performance. In practice, the IRS's passive beamforming is firstly challenged by the high computational complexity with a large size of the scattering elements. Similar to active beamforming of the RF transceivers, the IRS's passive beamforming also relies on the knowledge of channel information, which becomes more difficult to estimate for passive elements. Besides, the joint active and passive beamforming optimization is usually solved by convex approximations, leading to the convergence to a sub-optimal solution.

In this article, we focus on machine learning (ML) for performance maximization in IRS-assisted wireless networks with a large size of scattering elements. We aim at designing an efficient ML algorithm for joint beamforming that is robust against the channel dynamics and uncertain information. In Section~\ref{sec_irs}, we first provide an overview of IRS-assisted wireless networks, emphasizing the performance gains and the challenging issues. In Section~\ref{sec_mlsurvey}, we briefly review different applications of ML approaches in IRS-assisted wireless networks. Comparing to the optimization methods, ML approaches provide enhanced flexibility and robustness against uncertain information and imprecise modeling in a dynamic radio environment. The literature review also reveals some practical challenges to deploy ML approaches, mainly due to the requirement of a large amount of data samples for offline training or slow convergence in online learning. This observation motivates our work in Section~\ref{sec_oddpg}, where we design a novel optimization-driven ML framework to exploit both the efficiency of model-based optimization methods and the robustness of model-free ML approaches. The basic idea is to split the decision variables of a complex control problem into two parts. One part can be searched in the outer-loop ML approach, while the other part is optimized instantly by solving an approximate problem efficiently. Verified by numerical simulations, this method not only reduces the size of action space in the ML approach, but also speeds up the search for optimal solutions by exploiting the control problem's structural properties. Finally, some open issues are discussed in Section~\ref{sec_con}.

\section{IRS-assisted Wireless Systems}\label{sec_irs}

The IRS's reconfiguration relies on tunable chips embedded in the IRS structure~\cite{irsurvey}. As illustrated in Fig.~\ref{fig-roles}(a), each tunable chip is controlled by the IRS controller to adapt the phase shift of each scattering element. The IRS controller can communicate the reconfiguration settings with the external RF transceivers. In the following, we first discuss the IRS's different roles in a wireless network, and then review existing design problems for IRS-assisted wireless systems.

\subsection{Different Roles of IRS in Wireless Systems}

\subsubsection{Signal Reflector} The RF signal in the direct link between the RF transceivers can be combined coherently with its reflections at the receiver, or combined destructively to suppress information leakage to unintended receivers, as shown in Fig.~\ref{fig-roles}(b). Thus, it supports a higher data rate with reduced transmit power. The performance gain is shown to be proportional with the number of scattering elements~\cite{18pbf_rui2}. A large-size IRS can be divided into smaller groups and flexibly configured to create multiple reflections. Multiple IRSs can also work in collaboration to guide the reflected signals to reach ill-conditioned receivers.

\subsubsection{Signal Transmitter} By controlling the IRS's phase shifts in a time-varying manner, the outbound RF signals can exhibit different radiation patterns and thus carry useful information. This can be viewed as a generalization of the conventional backscatter communications~\cite{kim}, which modulate the information bits by varying the load impedance and thus changing the antenna's reflecting states. Whereas, the IRS-based backscatter communications have a higher flexibility as the IRS is able to generate more exotic reflection patterns for information communications, leading to a higher data rate and also a larger transmission distance.

\subsubsection{Signal Receiver} The IRS's reflecting elements can be viewed as individual receivers for multi-user data transmissions if the IRS is equipped with the signal processing unit. The IRS's reconfigurability can achieve an impressive capacity gain by suppressing the interference among different users. The large-size IRS can be divided into smaller units, processing individually the received signals from different users. The IRS can also be used as an array of sensors to estimate the position of the mobile devices based on the sensed variations of the RF power on the IRS.

\begin{figure*}[t]
  \centering
  \subfloat[IRS structure~\cite{irsurvey}]{\includegraphics[width=0.33\textwidth]{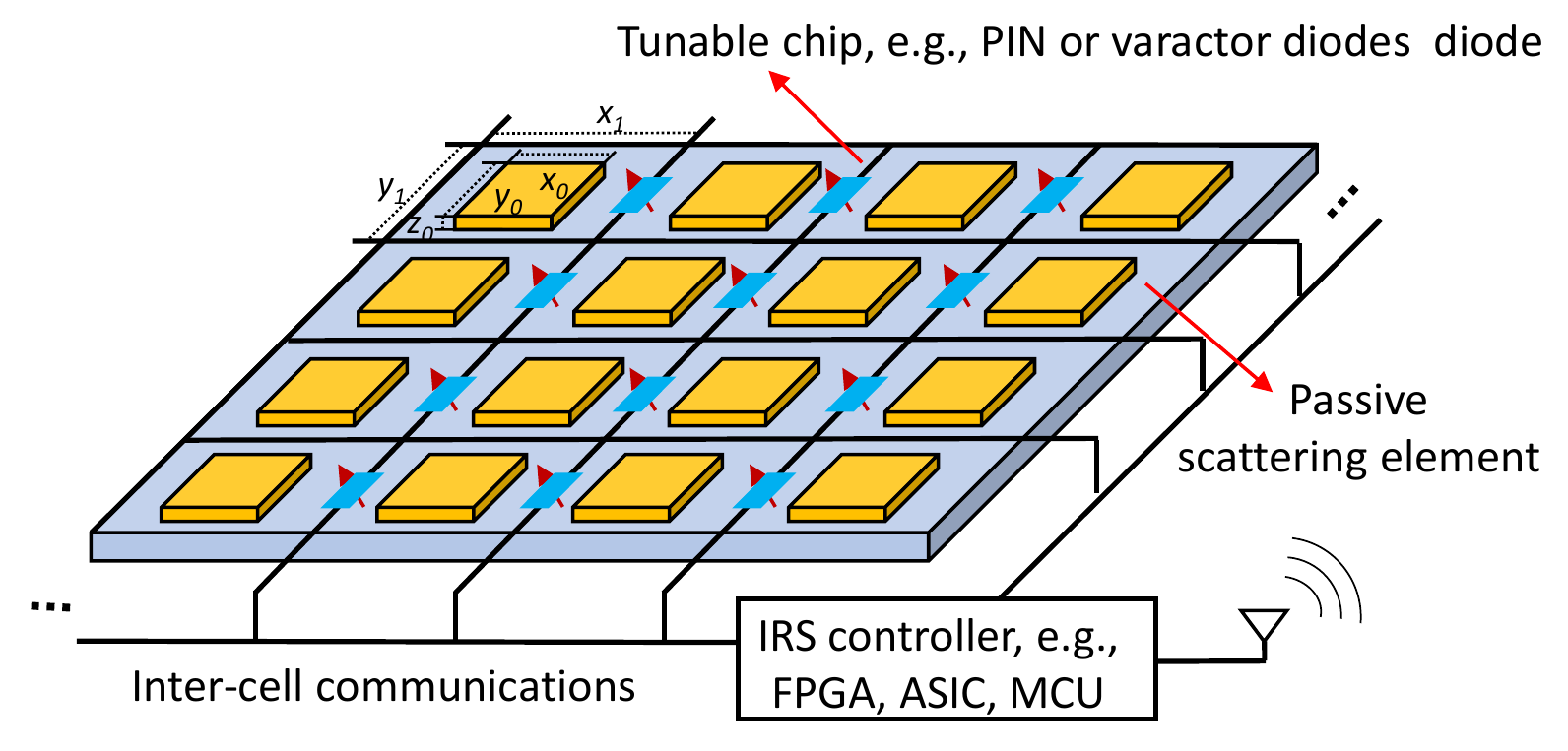}}
  \subfloat[Using IRS as reflector or transmitter]{\includegraphics[width=0.33\textwidth]{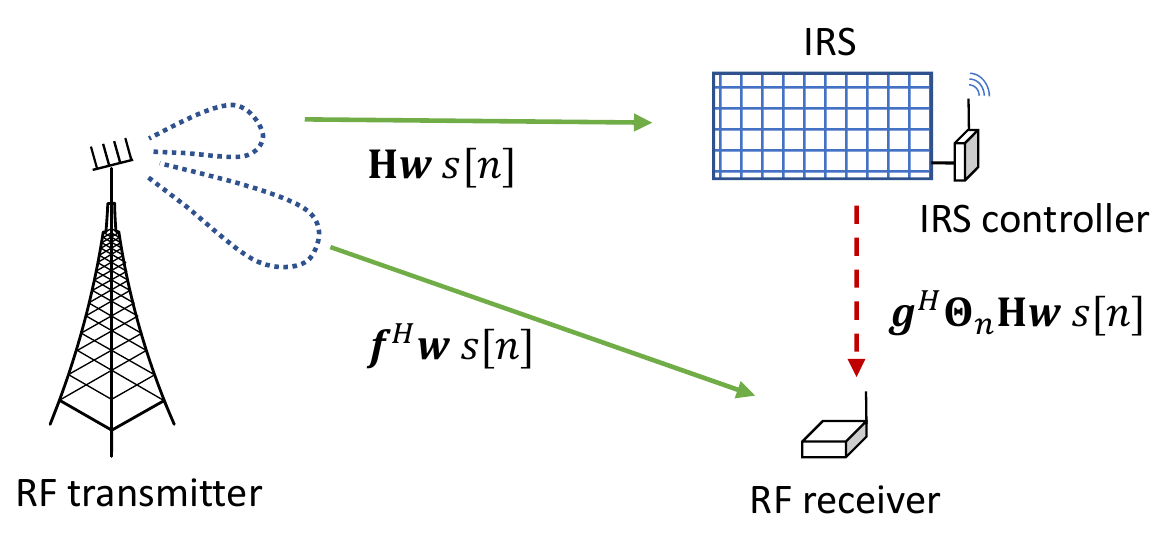}}
  \subfloat[Using IRS as receiver]{\includegraphics[width=0.33\textwidth]{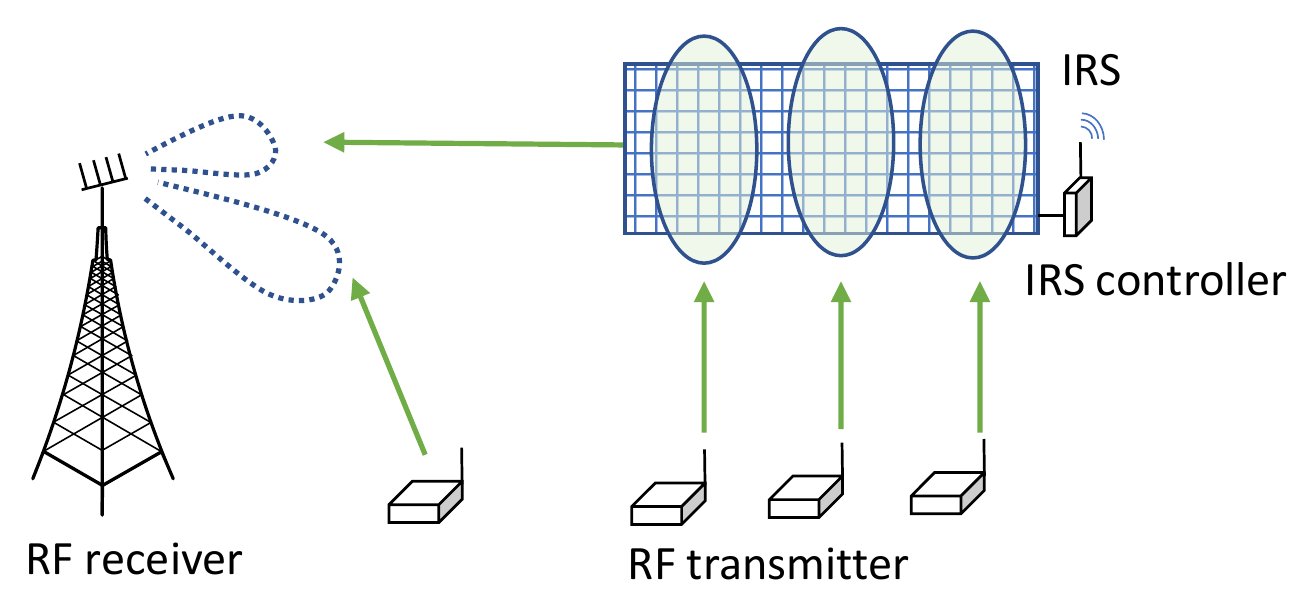}}
  \caption{(a) The IRS is composed by an array of passive scattering elements. (b) The IRS can reflect the incident RF signals towards intended directions by adjusting its passive beamforming matrix ${\bm \Theta_n}$. The IRS can also vary ${\bm \Theta_n}$ to create exotic radiation patterns that can carry different information bits from the IRS. (c) The IRS is viewed as a large array of receiver antennas for information reception and sensing.}\label{fig-roles}
\end{figure*}

\subsection{Potential Performance Gains of IRS-assisted Systems}

\subsubsection{Channel Capacity Gain}
The joint beamforming for capacity maximization with an ideal IRS can be decomposed in two sub-problems. The passive beamforming aims to maximize the equivalent channel gain combining the direct and the reflected links, while the active beamforming can be aligned with the equivalent channel. For a more practical case with a power-limited IRS~\cite{yuzetvt}, the active beamforming needs to transfer power to sustain the IRS's operations via RF energy harvesting. As the passive IRS has a cheaper implementation cost, we can use a large-size IRS to enhance the channel conditions and increase the received signal strength.

\subsubsection{Transmit Power Saving}
The enhanced channel condition also implies that the transmitter can save its transmit power while still maintaining the same level of quality provisioning to the users. This results in a more energy-efficient communication paradigm. The power scaling law revealed in~\cite{18pbf_rui2} shows that the base station's transmit power in an IRS-assisted system can be scaled down in the order of $1/N^2$ without compromising the receiver's performance, where $N$ denotes the number of the IRS's scattering elements. The power saving becomes more significant for wireless users far away from the base station. A similar power scaling law also holds for a practical IRS with low phase resolution.

\subsubsection{Improving Secrecy Rate} The IRS can be used to prevent wireless eavesdropping attacks by suppressing the information leakage to the illegitimate users. Simulation results in~\cite{Yu2019Enabling} reveal that it can be more effective to enhance secrecy rate and energy efficiency by deploying a large-scale IRS instead of increasing the size of antenna array at the active transmitter. Besides the prevention from eavesdropping by, the legitimate transmissions can also be hidden from being noticed or deciphered by the illegitimate user. This is achieved by leveraging the IRS to reshape undesirable propagation conditions and thus provide stronger protection from information leakage.

\subsection{Challenges for Active and Passive Beamforming}

The joint optimization in the literature is mainly solved by an alternating optimization (AO) method that decomposes the IRS's phase control and the RF radios' transmit beamforming into two sub-problems, e.g.,~\cite{yuzetvt} and~\cite{Yu2019Enabling}. In each sub-problem, semidefinite relaxation (SDR) is usually required to optimize the beamforming strategy by solving a semidefinite program (SDP) with high computational complexity. Though optimization-based methods provide predictable performances and even closed-form solutions, they generally suffer from the following difficulties in practice.

\subsubsection{High Computational Complexity} A larger size of the scattering elements ensures better flexibility in the IRS's phase control, even with a finite phase resolution or hardware impairments. Such a flexibility also comes with a cost. Typically, the IRS's phase control is formulated into an SDP, which has a high computational complexity as the size of IRS becomes large. A large-size IRS also implies huge training overhead and high power consumption for channel estimation.

\subsubsection{Unavailable Channel Information}
More sophisticated and agile protocols are required for channel estimation and information exchange, which should be completed within one coherence time to ensure valid channel information for beamforming control, especially in a dynamic environment. Moreover, a practical solution to the joint beamforming optimization should be robust against the error estimates or uncertainties in channel information.

\subsubsection{Imprecise System Modeling}
The joint beamforming problem is generally formulated based on a simplified system model, e.g.,~with perfect channel information, continuous and exact phase control. The problem reformulation and approximation further leads to a solution far from the optimum, which implies unpredictable performance in practice. We expect that the optimization methods only provide a lower or upper bound on the original problem.

The above challenges motivate us to use model-free ML approaches to solve the joint beamforming problem for IRS-assisted systems. In the sequel, we first review the applications of ML approaches and then analyze the current limitations, which motivate our design for a novel optimization-driven ML framework in Section~\ref{sec_oddpg}.

\section{Machine Learning for IRS-assisted Wireless Systems}\label{sec_mlsurvey}
ML approaches include supervised and unsupervised learning, depending on the availability of labeled samples in the training data. The learning performance can be improved by leveraging deep neural networks (DNNs) to extract hidden features of the training data. The DNN weight parameters can be adjusted iteratively based on the labeled data for supervised learning or unlabeled data for unsupervised learning. Reinforcement learning (RL) makes decisions by continuously interacting with the uncertain environment, i.e.,~the decision-making agent takes an action and then receives an immediate reward based on the observation of the environment. The RL framework becomes unstable when the state and action spaces are large in complex systems. Deep reinforcement learning (DRL) overcomes this difficulty by using DNNs to approximate different components of the RL framework. In particular, the deep $Q$-network (DQN) algorithm uses DNNs to approximate the value function. The deep deterministic policy gradient (DDPG) algorithm uses two set of DNNs, namely, the critic and actor networks, to estimate the value function and the optimal policy, respectively.

\subsection{Channel Estimation and Signal Detection}
The channel estimation can be performed in a training period by sending a sequence of known pilot at the transmitter and then estimating the channel information based on the observed channel response at the receiver. The input pilot and the expected channel response can be viewed as the labeled data for supervised learning. A large-size IRS will generate a huge dataset during the channel training, which can be flexibly handled by DNNs to enhance the training performance. For example, the convolutional neural network (CNN) is employed in~\cite{dl-mmwave} to estimate both direct and cascaded channels for an IRS-assisted system, based on simulated input signals and the expected output channel vectors. The well-trained CNN is then used to predict the real-time channel conditions. Similarly, unsupervised learning can be used for signal detection in an IRS-assisted system based on simulated channel conditions and phase variances~\cite{dl-detection}. The DNN weights are trained to minimize the difference between the received and the transmitted signal vectors.

\begin{table}[t]
\caption{ML approaches in IRS-assisted Wireless Networks}
\centering
\resizebox{1\textwidth}{!}{\begin{tabular}{ p{6mm}  p{10mm}  p{28mm}  p{25mm}  p{55mm} }
\hline
REF & Model & ML approach   & Objective   & Contributions  \\\hline
\cite{dl-mmwave} & MIMO & Supervised, CNN & Channel estimation   & Less error and more robust performance  \\\hline
\cite{dl-detection}& MISO & Unsupervised, DNN & Signal detection   & Near-optimal BER performance \\\hline
\cite{dl-asu} & OFDM & Supervised, DNN & Data rate  & Achieve the upper bound with perfect channel information   \\\hline
\cite{unsupervised} & MISO & Unsupervised, DNN &  Received SNR  & Comparable performance with the SDR method \\\hline
\cite{walid} & MISO  & DQN & Energy efficiency  & Energy efficiency increases up to 77.3\% when $N$ is 25\\\hline
\cite{drl-miso} & MISO & DDPG & Received SNR  & Achieve close-to-optimal SNR with low time consumption \\\hline
\cite{yuan-ddpg} & MISO & DDPG & Sum rate &  Achieve comparable performance with optimization methods\\\hline
\cite{drl-dusit} & MISO & DQN & Secrecy rate  & Improve secrecy rate and QoS  \\\hline
\end{tabular}}
\end{table}

\subsection{Machine Learning for Passive Beamforming}

DNNs can be trained to recall a high-dimensional mapping from the environmental feature (e.g.,~channel response and the receiver's location) to the optimal passive beamforming. The receiver's location can be mapped to the IRS's optimal phase configuration that maximizes the received signal strength~\cite{dl-asu}. At each location, the optimal phase vector can be determined by an exhaustive search method, which serves as the labeled data sample. The well-trained DNN is then used for online prediction of the IRS's optimal phase vector given the user's position. Unsupervised learning is also used for optimizing the passive beamforming to maximize the signal-to-noise ratio (SNR) at the receiver~\cite{unsupervised}. This can be implemented by training the DNN to minimize the loss function, defined as the negative of the averaged SNR at the receiver. DRL becomes more flexible to solve complicated problems with uncertain and dynamic conditions. It requires a reformulation into Markov decision process (MDP) with properly defined system state, action set, and the reward function. As such, the DQN algorithm can be used for the IRS to update its action based on the observed channel conditions and the receiver's feedback~\cite{walid}. The continuous phase vector can be directly optimized by using the DDPG algorithm to maximize the received SNR~\cite{drl-miso} and the sum rate~\cite{yuan-ddpg}, which reveal that the DDPG algorithm can achieve comparable performance as that of the conventional optimization-based algorithms with significantly reduced time consumption.

\subsection{Anti-Jamming and Secure Communications}
The RL and DRL approaches can also be used to enhance physical layer security of IRS-assisted systems against eavesdroppers or active jammers. To prevent eavesdropping by, the joint beamforming aims at maximizing the secrecy rate, i.e.,~the difference between the data rate to the legitimate users and the information leakage to the eavesdroppers. DRL solves this problem by formulating the reward function as the difference between the secrecy rate and a penalty term, accounting for the quality of service (QoS) at the receivers~\cite{drl-dusit}.


Compared to optimization-based methods, the ML approaches in IRS-assisted systems demonstrate more flexibility and robustness against uncertain information, imprecise modeling, and dynamic environment. However, we observe that their practical implementations are still challenging, mainly due to the requirement for a large amount of data samples in offline training or slow convergence in online learning. In particular, the DNN training in~\cite{dl-mmwave,dl-detection,dl-asu,unsupervised} have to either collect or randomly generate a sufficiently large dataset via simulations. The simulated data is typically based on simplified models, which may introduce systematic bias for online prediction. Though RL and DRL methods can learn to make decisions from scratch~\cite{walid,drl-miso,yuan-ddpg,drl-dusit}, they are subject to slow convergence via the interaction with the radio environment.

\section{Optimization-driven DRL Framework for Joint Beamforming}\label{sec_oddpg}

In this part, we aim to improve the learning efficiency by proposing a new learning framework for IRS-assisted systems, which exploits the efficiency of model-based optimization methods and the robustness of model-free ML approaches. A recent work in~\cite{model-based} proposed a similar concept of model-aided Artificial Intelligence (AI) in wireless systems, which leverages the model-based optimization to create a large set of \emph{offline} training data for optimizing or refining the DNN models, e.g.,~\cite{dl-asu} and~\cite{unsupervised}. This framework is verified to work well for some special cases with the availability of either an accurate model or a tractable optimization solution. However, it can become inflexible in dynamic conditions when the system states are uncertain and evolving correlated over time. Different from~\cite{model-based}, we use the DRL approaches to build a robust outer-loop learning framework that is tolerable to uncertain information and system dynamics, while using the inner-loop optimization methods during the \emph{online} learning phase to fast track the control variables by solving approximate optimization problems efficiently. As such, it can be applied to more complex wireless systems with both inaccurate models and intractable solutions.

\subsection{IRS-Assisted MISO System with Channel Uncertainties}

We consider a generic IRS-assisted multi-input single-output (MISO) downlink system, where the information transmissions from a multi-antenna access point (AP) to the receivers are assisted by the IRS with $N$ reflecting elements, similar to the system model illustrated in Fig.~\ref{fig-roles}(b). A few system assumptions are listed as follows:
\begin{itemize}
  \item The IRS can set a continuous phase shift and a flexible magnitude of reflection, i.e.,~the power-splitting (PS) ratio, to reflect the incident RF signals. The extension to discrete phase shift and discrete PS ratio is straightforward by using quantization projection.
  \item The IRS is self-sustainable by harvesting RF energy. By setting the PS ratio, a part of the incident signal power is reflected to the receiver, while the other part is absorbed by the energy harvester to fulfill the IRS's power demand.
  \item The channel estimation, beamforming control, and data transmission can be completed within each coherence time. Hence, a time-efficient ML approach ensures its applicability with dynamic channel conditions, e.g.,~in both outdoor and indoor environment.
  \item The channel estimations are inevitably subject to errors due to the use of passive elements at the IRS. The average estimates can be known by channel measurements, while the error estimates are randomly distributed within a convex and bounded set.
\end{itemize}

We aim to minimize the AP's RF power for information transmission by the joint beamforming optimization, subject to the IRS's power constraint and the receiver's SNR requirement. The optimization problem is firstly challenged by the non-convex coupling between the active and passive beamforming. The other difficulty comes from the uncertain channel conditions, leading to either a stochastic or worst-case robust reformulation with high computational complexity.

\subsection{Optimization-driven DRL Framework}
In the sequel, we solve the above transmit power minimization problem by using the DRL framework that can well handle inaccuracies in system modeling and the uncertainties in channel conditions. We first analyze the drawbacks of the conventional DRL approaches and then propose a novel DRL approach with enhanced learning efficiency.

\begin{figure*}[t]
  \centering
  \subfloat[Model-free DQN algorithm]{\includegraphics[width=0.45\textwidth]{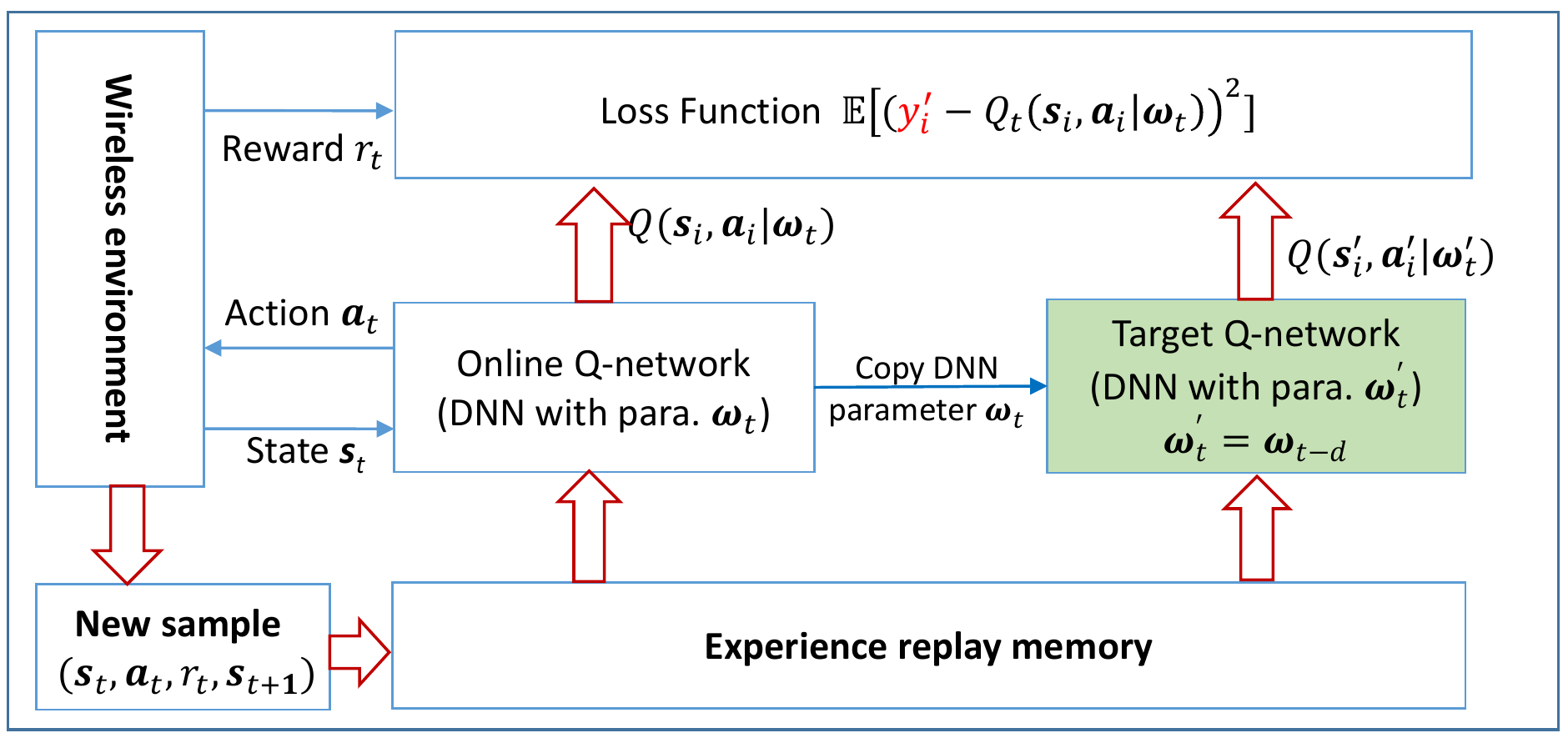}}
  \subfloat[Model-free DDPG algorithm]{\includegraphics[width=0.45\textwidth]{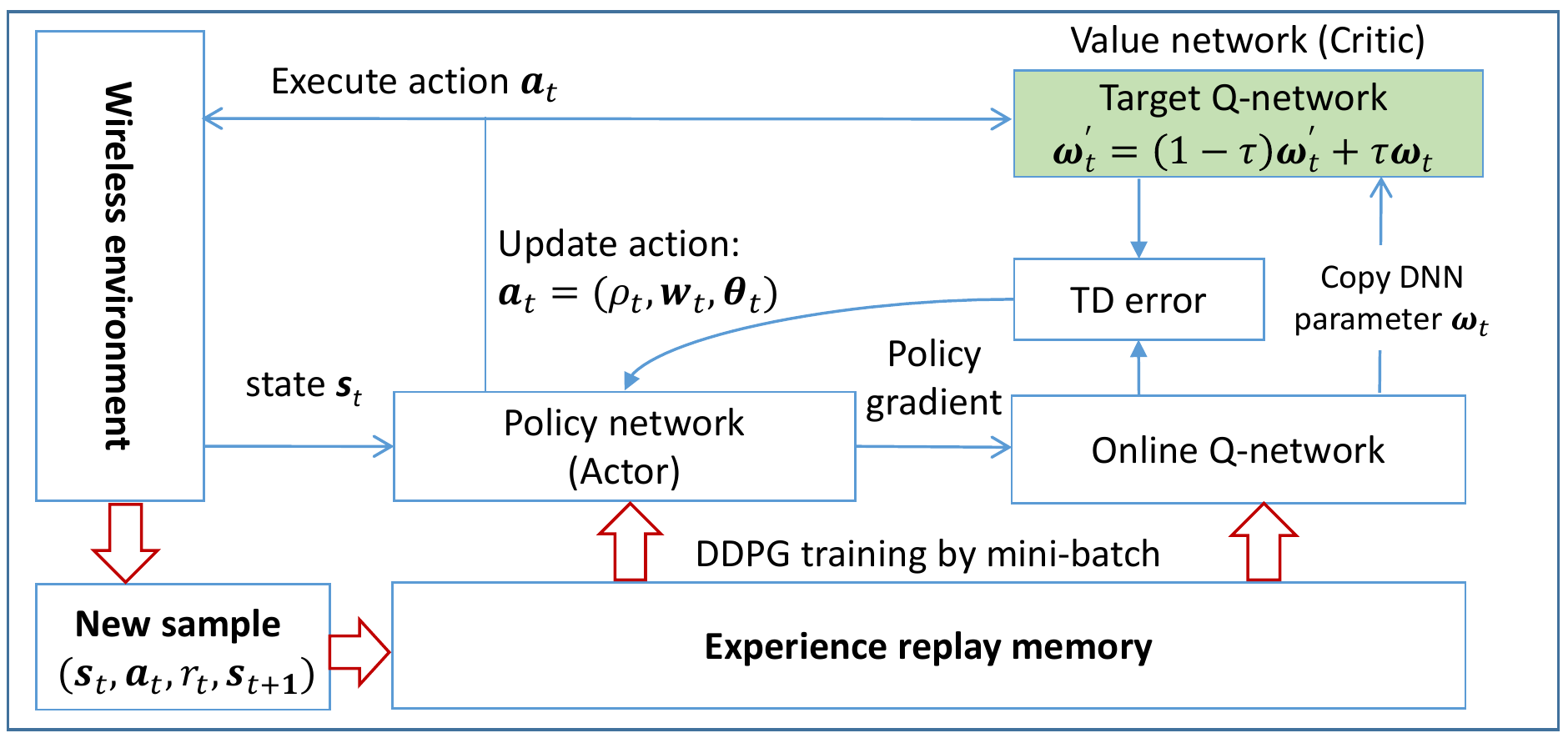}}\\
  \subfloat[Optimization-driven DQN algorithm]{\includegraphics[width=0.45\textwidth]{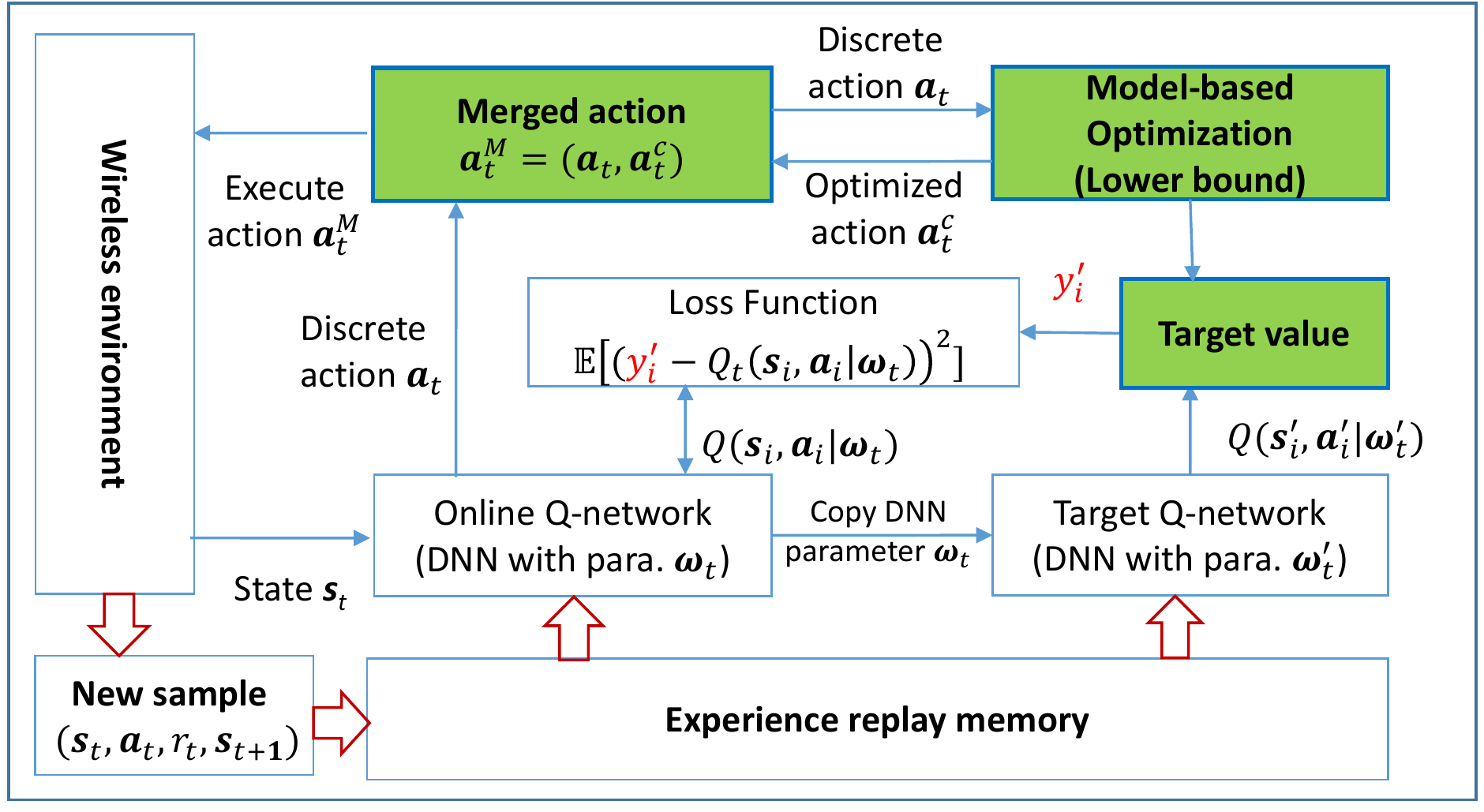}}
  \subfloat[Optimization-driven DDPG algorithm]{\includegraphics[width=0.45\textwidth]{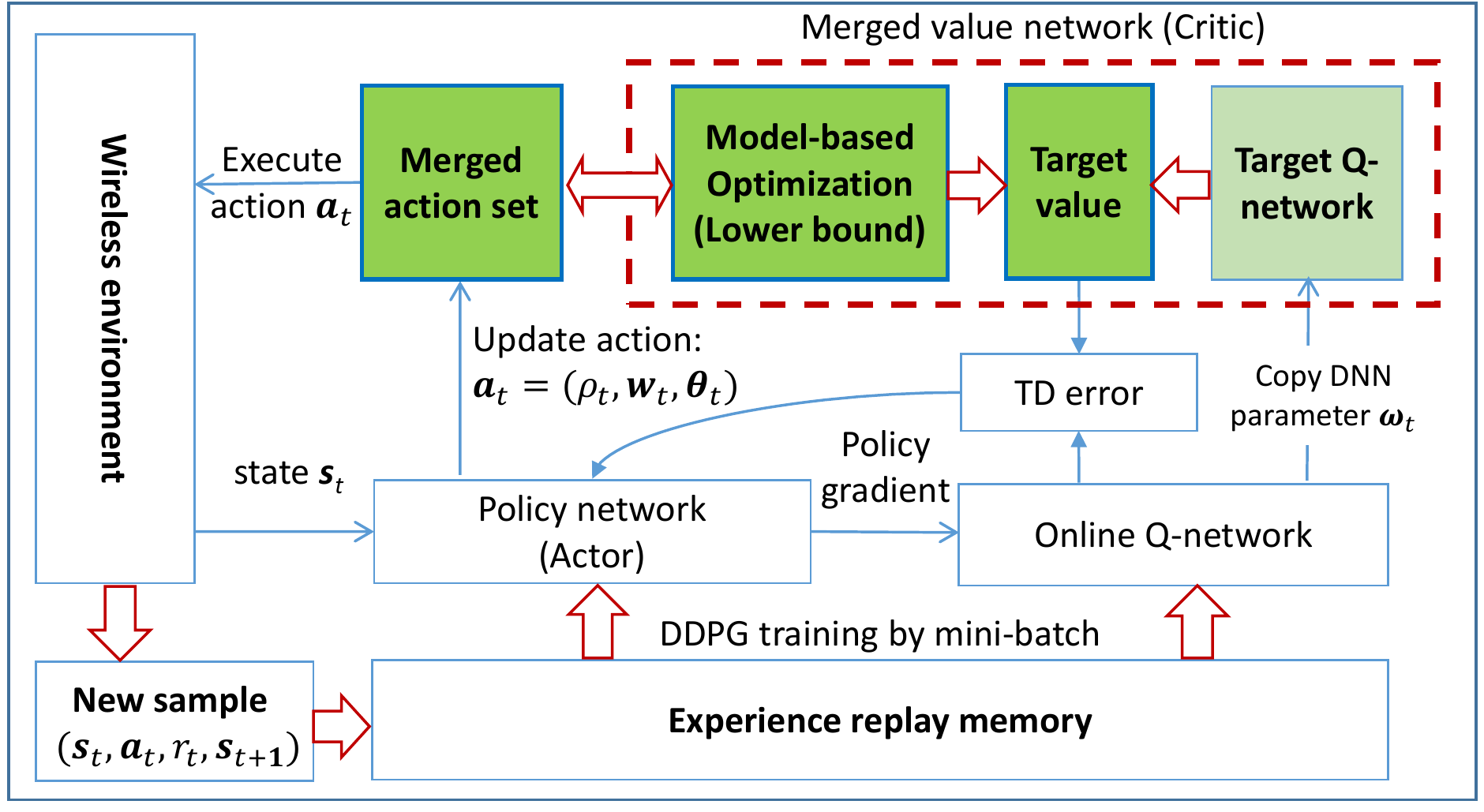}}\\
  \subfloat[The application of optimization-driven learning framework in IRS-assisted systems]{\includegraphics[width=0.9\textwidth]{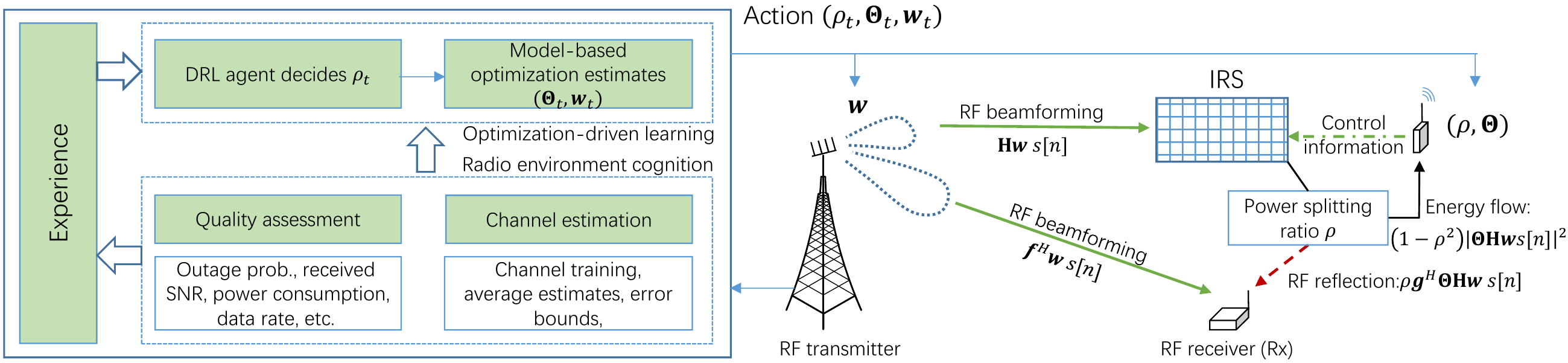}}
  \caption{(a)-(d) The comparison of model-free and optimization-driven DQN/DDPG algorithm. $y_t'$ denotes the lower bound of the target $Q$-value determined by the optimization module. (e) The general optimization-driven DRL framework for joint beamforming optimization in the IRS-assisted MISO downlink system. There can be also multiple distributed IRSs that jointly assist the MISO downlink communications. In this case, coordination among IRSs is required for the estimation of the AP-IRS and AP-IRS-User channels.}\label{fig-drl}
\end{figure*}

\subsubsection{Conventional Model-free DQN and DDPG Algorithms}

The conventional DQN algorithm relies on the use of experience replay and target $Q$-network to stabilize the learning performance. The experience replay mechanism randomly selects a mini-batch from a buffer of historical samples to train the DNN. As illustrated in Fig.~\ref{fig-drl}(a), the DNN training updates the DNN parameters of the online $Q$-network by minimizing the temporal-difference (TD) error, i.e., the mean-squared difference between the online and the target $Q$-values. To stabilize the learning performance, the DQN algorithm estimates the target $Q$-value by using a separate DNN, namely, the target $Q$-network, whose parameters are delayed copies of the online $Q$-network after every a few decision epoches. The same idea also applies to the DDPG algorithm. As shown in Fig.~\ref{fig-drl}(b), the estimation of $Q$-value in the critic is accompanied by a separate target $Q$-network, whose parameters are also evolving from the online $Q$-network.

Though the target $Q$-network in DQN or DDPG algorithm stabilizes the learning performance, the strong coupling between the online and target $Q$-networks may lead to a slow learning rate and reduced reward performance. Firstly, both $Q$-networks can be randomly initialized and could be far from their optimum in the early stage of learning. This may mislead the learning process and require a large training dataset, i.e.,~historical transition samples, to ensure the learning toward the right direction. As such, the model-free DQN and DDPG algorithms practically require a long warm-up period to train the online and target $Q$-networks. Secondly, it is problematic to configure the parameter copying from the online $Q$-network to the target $Q$-network. As shown in Fig.~\ref{fig-drl}(b), a small averaging parameter $\tau$ for the DDPG algorithm can stabilize but also slow down the learning process, while a large $\tau$ implies strong correlation between the two $Q$-networks, resulting in performance fluctuations and even divergence. 

\subsubsection{Merging Model-free and Model-based Target $Q$-Values}

To improve learning efficiency, we design the optimization-driven DRL framework that integrates the model-based optimization into the model-free DRL framework. We aim to stabilize and speed up the learning process by estimating the target $Q$-value in a better-informed and independent way. The motivation for the proposed framework stems from the following observations: 1) The control variables of a complex problem are usually high dimensional and untractable by classical optimization methods. 2) However, given a part of the decision variables, the other part can be easily optimized in an approximate problem by exploiting their physical connections. Therefore, the main design principle is to split the control variables into two parts. One part with reduced search space will be obtained in the outer-loop DRL approach, e.g.,~the DQN or DDPG algorithm, while the other part is optimized instantly and efficiently given the outer-loop control variables.
\begin{itemize}
  \item Optimization-driven DQN: For a mixed problem with both discrete and continuous decision variables, we can split the action vector into two parts, i.e.,~the discrete and continuous variables $({\bf a}_t, {\bf a}_t^c)$ as shown in Fig.~\ref{fig-drl}(c). The discrete action can be learnt from the conventional DQN algorithm. Given the discrete action, we can resort to the optimization method to solve efficiently the continuous actions in a simplified version of the original problem, which also provides a lower bound on the target $Q$-value. The optimized action can be combined with the discrete action and then executed in the radio environment.
  \item Optimization-driven DDPG: For a high-dimensional control problem, the divide and conquer principle can be similarly applied to improve the learning efficiency of the conventional DDPG algorithm. As illustrated in Fig.~\ref{fig-drl}(d), when the DDPG algorithm generates a part of the action, an optimization module can solve the other part of the action directly and provide a lower bound on the original problem. The partition of the action vector should be properly designed such that the model-based optimization can be solved efficiently.
\end{itemize}

We envision that the model-based optimization can provide a better-informed target compared to that generated by the target $Q$-network, especially in the early stage of learning. Moreover, the optimization-driven target is independent of the online $Q$-network. It can be more stable than the output from the target $Q$-network. Such a decoupling between the online and target $Q$-networks can reduce the performance fluctuation and thus stabilize the learning faster.

\subsubsection{Optimization-driven DDPG for Joint Beamforming}
The joint beamforming optimization for the IRS-assisted MISO downlink system includes the AP's active beamforming vector ${\bf w}$, the IRS's PS ratio $\rho$ and the phase vector ${\bm \theta}$, as illustrated in Fig.~\ref{fig-drl}(e). We employ the optimization-driven DDPG algorithm to search for the optimal action ${\bf a}_t = (\rho_t, {\bf w}_t, {\bm \theta}_t)$ in each decision epoch, which can be divided into two parts, i.e.,~a scalar $\rho_t$ and two vectors $({\bf w}_t, {\bm \theta}_t)$. Given the PS ratio $\rho_t$, we can easily determine a feasible phase vector ${\bm \theta}_t$ and then solve the optimal active beamforming ${\bf w}_t$ efficiently in a convex optimization problem. This implies that we can construct an optimization module to update $({\bf w}_t,{\bm \theta}_t)$ and use the DDPG algorithm to search for $\rho_t$~\cite{jiaye}. This not only reduces the search space of the DDPG algorithm, but also speeds up the search for $({\bf w}_t,{\bm \theta}_t)$, compared to the conventional model-free DDPG algorithm. As shown in Fig.~\ref{fig-drl}(d), the actor and critic networks of the DDPG algorithm firstly generate the action and value estimates independently. Then, we fix $\rho_t$ in the action and feed it into the model-based optimization module, which outputs the optimized solution $({\bf w}'_t, {\bm \theta}'_t)$ and also evaluates a lower bound $y_t'$ on the target $Q$-value. If the optimization-driven target is larger than the output of the target $Q$-network, we can use $y_t'$ with a higher probability as the target $Q$-value for DNN training, meanwhile update the action as~$(\rho_t,{\bf w}'_t, {\bm \theta}'_t)$. When the learning becomes more stable, the optimization-driven target may be smaller and then we can follow the actor's decision $(\rho_t,{\bf w}_t, {\bm \theta}_t)$.

\subsection{Numerical Evaluation}

We consider the system model as in Fig.~\ref{fig-drl}(e). The AP-User distance in meters is set as $d_{AP,User}= 20$. The vertical distance from the IRS to the AP-User line segment is set to $d_{IRS,User}=5$. The IRS can move away from the AP to the receiver. The path loss at the reference point is given by $L_0 = 30$~dB and the path loss exponent equals 3.5, similar to~\cite{18pbf_rui2}. The energy harvesting efficiency is set as $\eta = 0.5$. The noise power is $-80$ dBm. We aim to minimize the AP's transmit power by the proposed optimization-driven DDPG algorithm. The reward function is defined as the ratio between the successfully transmitted data and the AP's total power consumption. Given $\rho_t$ to the optimization module, the passive beamforming ${\bm \theta}_t$ is designed by heuristic to enhance the direct link while the active beamforming ${\bf w}_t$ is optimized optimally by solving an SDP efficiently, following a similar approach as that in~\cite{jiaye}.

\begin{figure}[t]
  \centering
  \subfloat[AP's transmit power]{\includegraphics[width=0.45\textwidth]{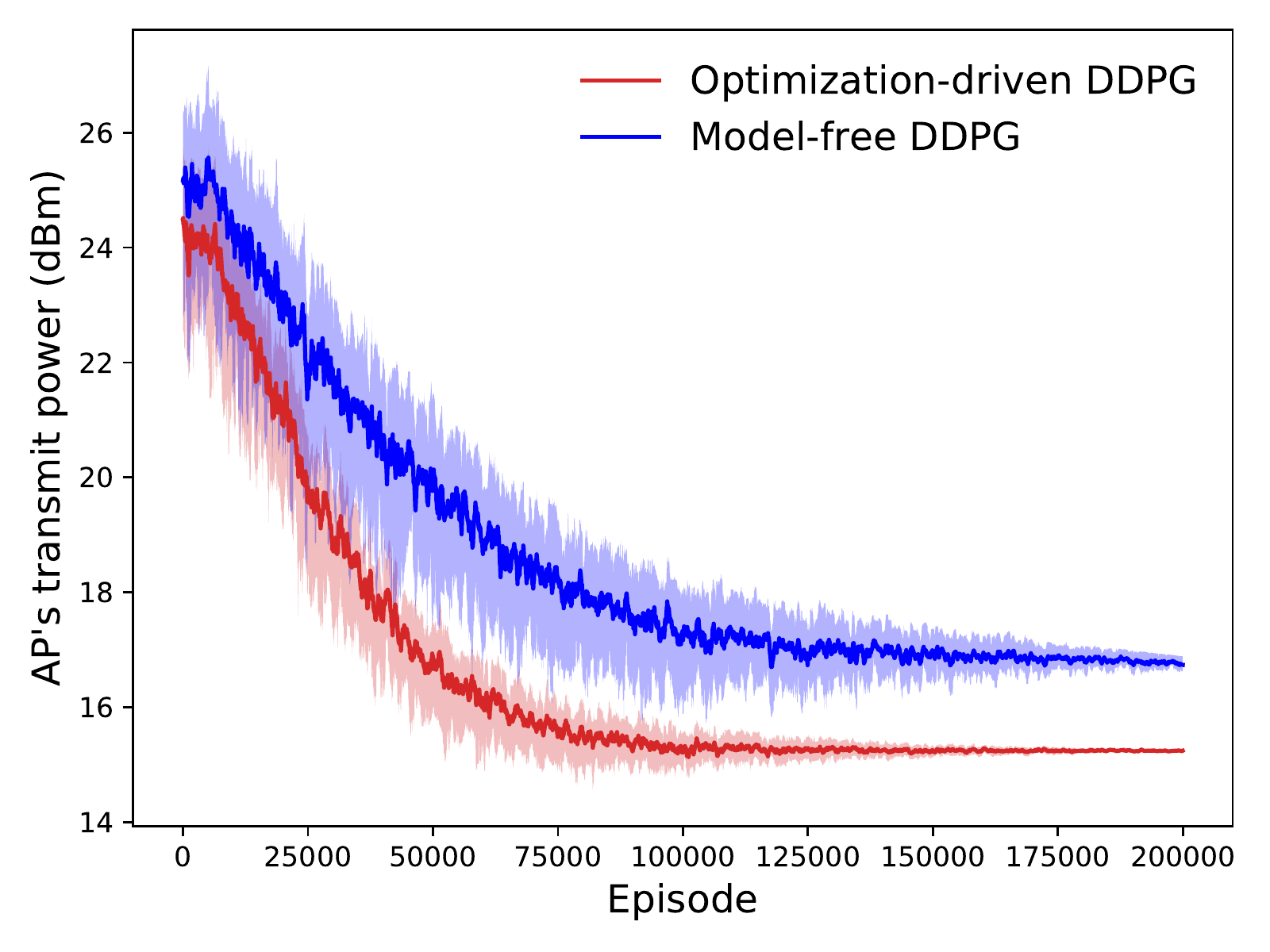}}
  \subfloat[IRS's PS ratio]{\includegraphics[width=0.45\textwidth]{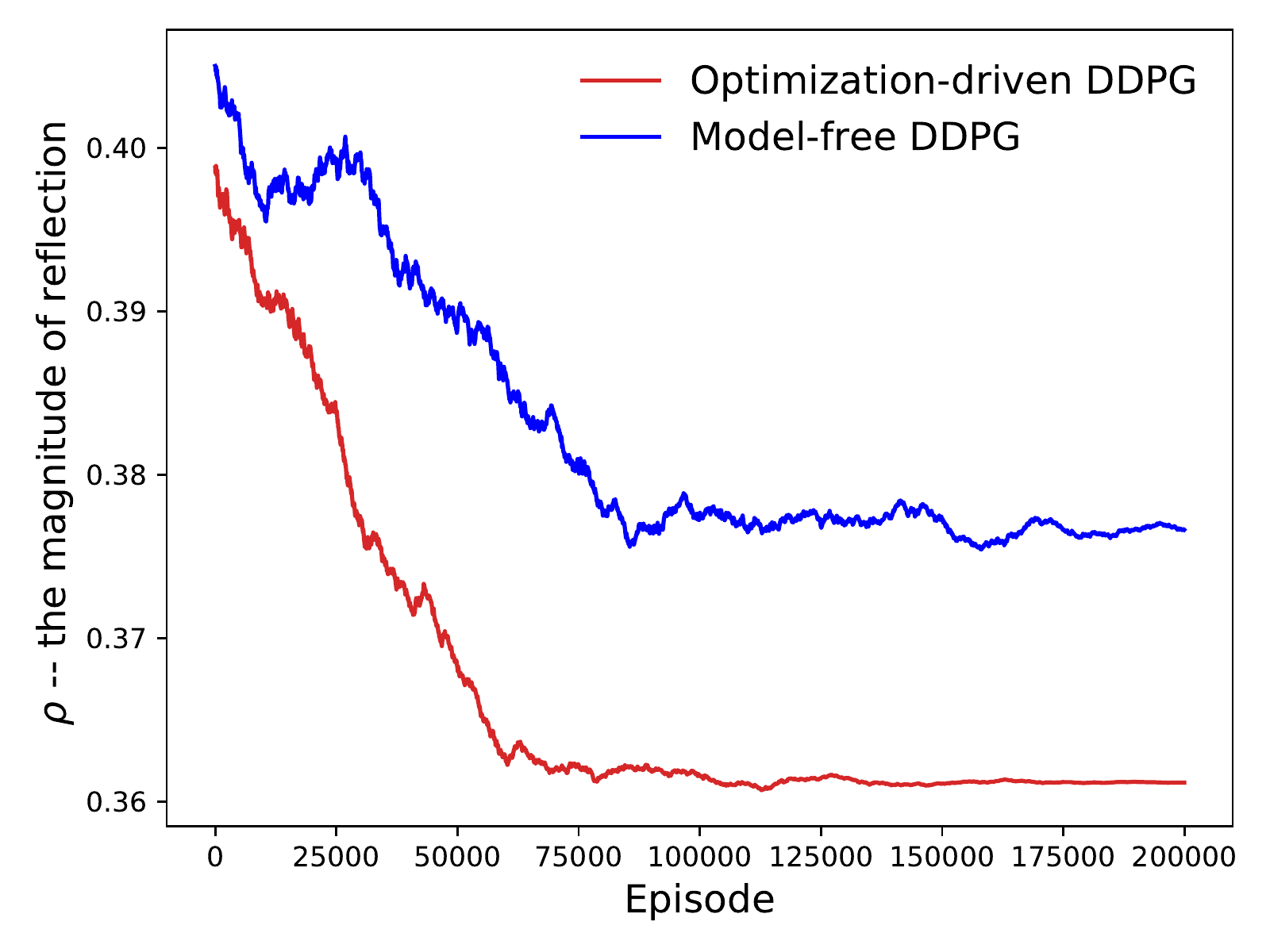}}
  \caption{The AP's transmit power and the IRS' PS ratio in two DDPG algorithms. The solid line denotes the median of 50 repetitions and the shaded regions in different colors cover 10th to 90th percentiles. The AP's and user's locations are set as $d_{AP,User} = 20$, $d_{AP,IRS}=10$, $d_{IRS,User}=5$.}\label{fig:conv}
\end{figure}

\begin{figure}[t]
  \centering
  \subfloat[Power decreasing]{\includegraphics[width=0.45\textwidth]{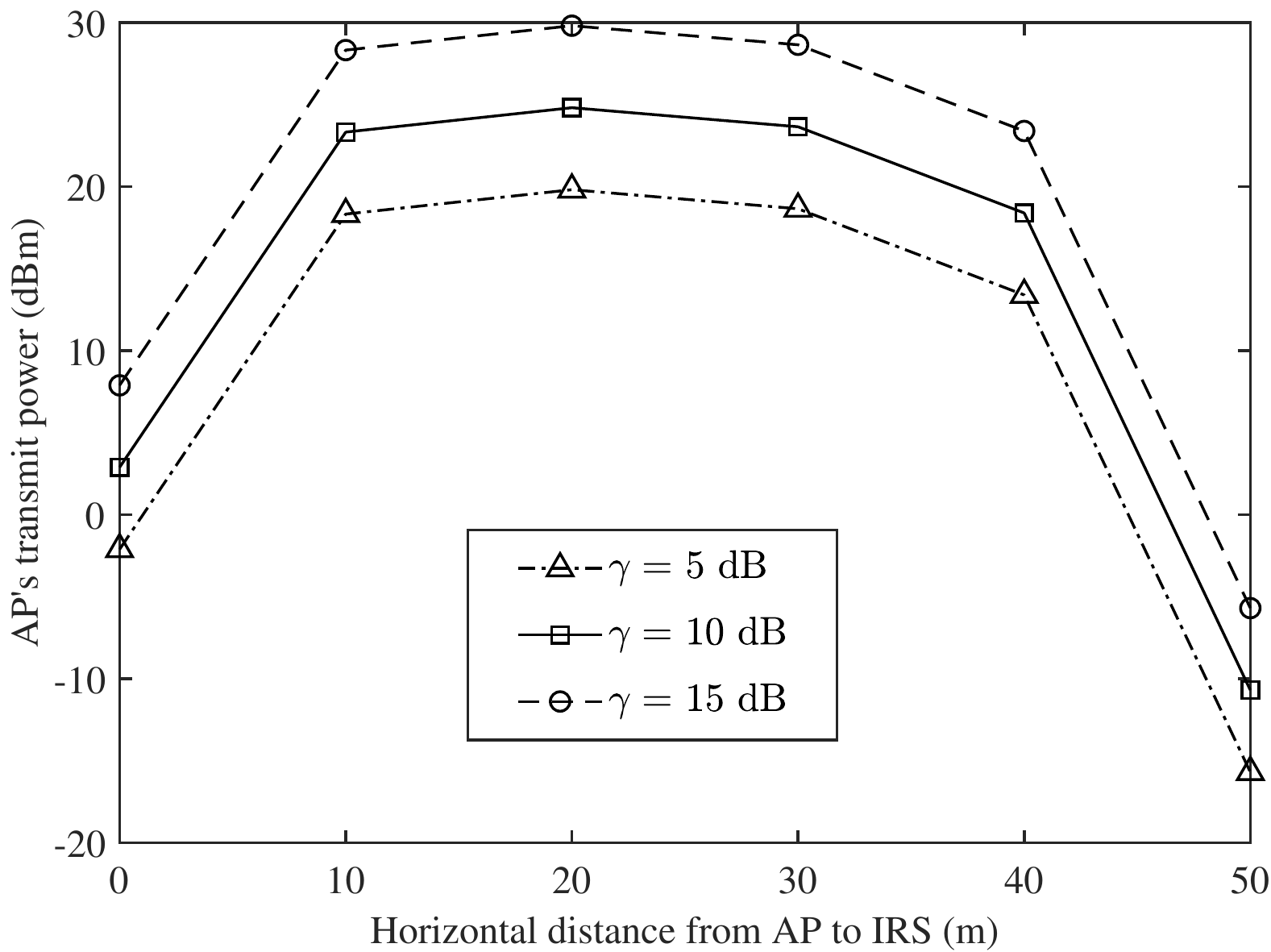}}
  \subfloat[Power increasing]{\includegraphics[width=0.45\textwidth]{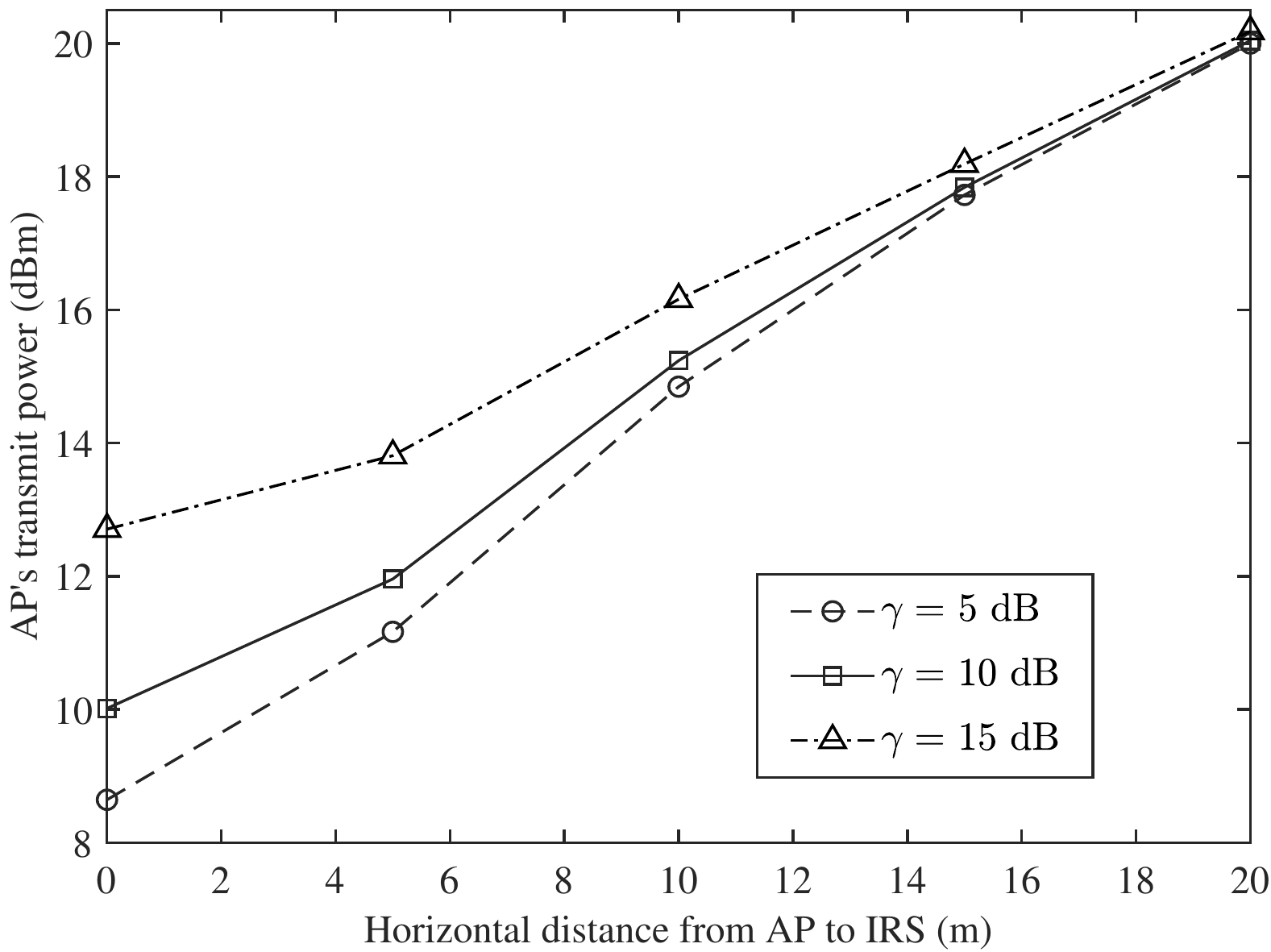}}
  \caption{The change of the AP's transmit power in the optimization-driven DDGP algorithm as the IRS moves from the AP to the receiver: a) The IRS has no power demand in the ideal case. b) The IRS has the constant power demand 20~$\mu W$, which has to be fulfilled by RF energy harvesting.}\label{fig:distance}
\end{figure}

\begin{figure}[t]
  \centering
  \subfloat[More stable learning]{\includegraphics[width=0.45\textwidth]{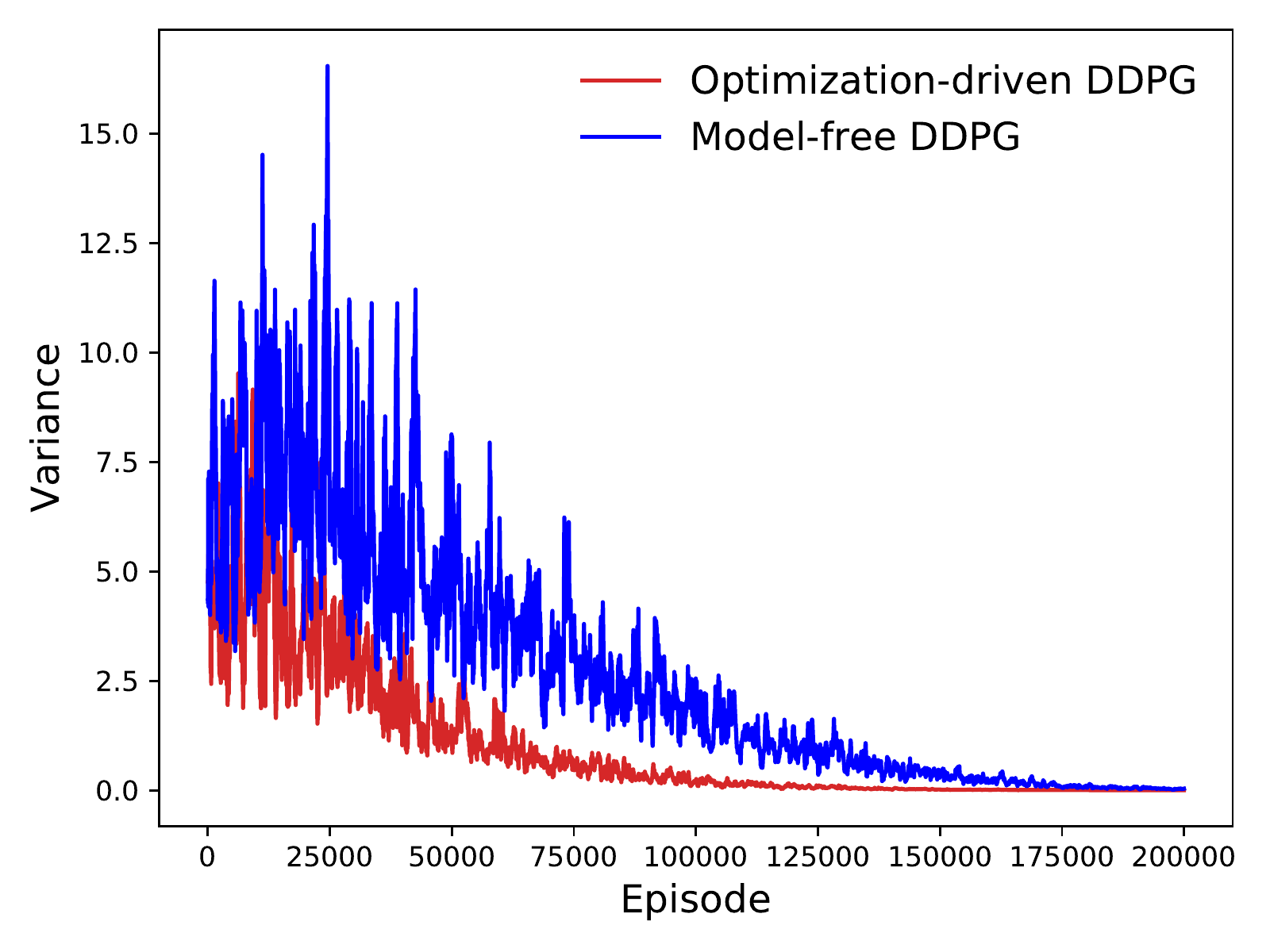}}
  \subfloat[Higher time efficiency]{\includegraphics[width=0.45\textwidth]{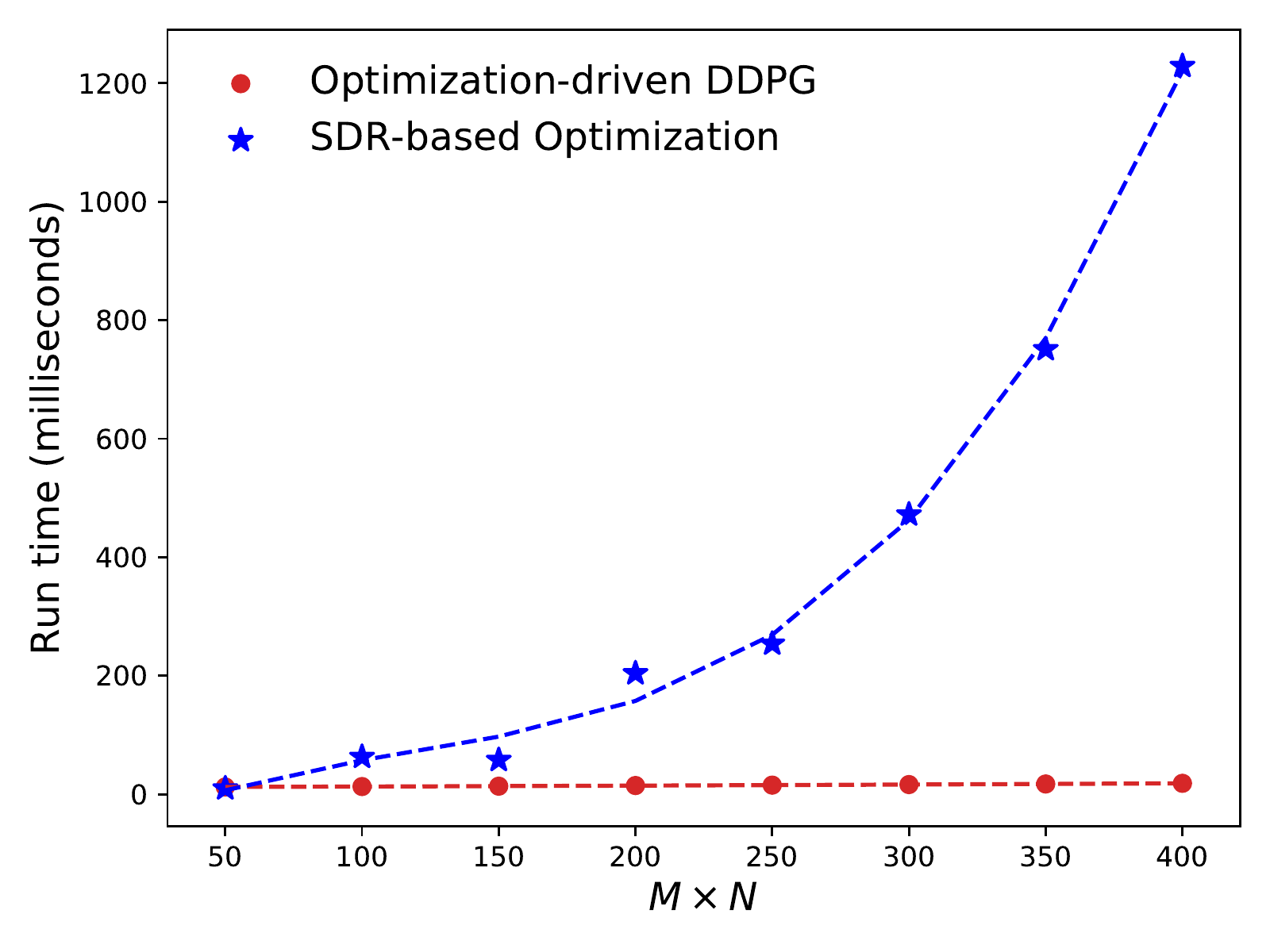}}
  \caption{(a) The optimization-driven DDPG algorithm achieves more stable learning and faster convergence than the conventional model-free DDPG algorithm, (b) it also achieves better scalability as the problem size $M\times N$ increases, where $N$ and $M$ denote the numbers of the IRS's passive elements and the AP's active antennas, respectively. The dotted lines represent the polynomial curve fitting.}\label{fig:stable}
\end{figure}

\subsubsection{Faster Convergence}
Figure~\ref{fig:conv} demonstrates the AP's transmit power and the IRS's PS ratio $\rho$ in the optimization-driven DDPG algorithm. We also compare them to the conventional DDPG algorithm, denoted as model-free DDPG. It is clear that the optimization-driven DDPG converges faster than the model-free DDPG and achieves a significant performance improvement in terms of the AP's transmit power. The reason is that the optimization-driven DDPG uses a better-informed estimation for the target value to guide its search for the optimal policy. Figure~\ref{fig:distance} shows the AP's transmit power when the IRS moves away from the AP to the receiver. In the ideal case when the IRS has no power demand, the AP can reduce its transmit power significantly as the IRS is closer to the receiver, as shown in Fig.~\ref{fig:distance}(a). This corroborates the observations in~\cite{18pbf_rui2}. However, given the IRS's power demand, the AP needs to increase its transmit power as the IRS moves away from the AP, as shown in Fig.~\ref{fig:distance}(b). This indicates that the IRS's power demand becomes the performance bottleneck when it is deployed far from the AP.

\subsubsection{More Stable Learning} In Fig.~\ref{fig:conv}(a), we also observe that the shaded area of the optimization-driven DDPG is smaller than that of the model-free DDPG. This implies that a more stable learning performance in the optimization-driven DDPG. To verify this, we characterize the performance fluctuation by the variance of reward. As shown in Fig.~\ref{fig:stable}(a), the reward performance of the optimization-driven DDPG has much smaller variance compared to that of the model-free DDPG. The convergence in the optimization-driven DDPG also comes earlier as the variance approximates to zero, which corroborates the observations in Fig.~\ref{fig:conv}.

\subsubsection{Better Scalability}
To show the scalability of the optimization-driven DDPG approach, we compare its average run time in each decision epoch with the conventional SDR-based optimization method, e.g.,~\cite{18pbf_rui2} and~\cite{yuzetvt}, which has an increasing computational complexity in terms of the problem size. As shown in Fig.~\ref{fig:stable}(b), the optimization-driven DDPG algorithm has nearly a constant run time. Such a low complexity makes it very suitable for practical deployment, especially with a large-size IRS and a large number of active antennas.

\section{Open Research Issues}\label{sec_open}

Though the optimization-driven DRL has numerically shown significant performance gain for the joint beamforming optimization, there still exist some challenges and open issues as follows:

\emph{1) Control Variable Division:} The optimization-driven DRL framework divides the control variables into outer-loop learning and inner-loop optimization. This division imposes the performance tradeoff between the computational complexity of optimization methods and the time efficiency of learning methods. The optimal variable division can be characterized in the future.

\emph{2) Convergence Property:} The outer-loop learning has a reduced action space, which implies an increased learning speed, however with the potential cost of performance loss. Though numerical results show preferable performance gain, a formal proof for convergence or performance guarantee can be further investigated.

\emph{3) Adaptive Integration:} Numerical results show that the optimization method improves the learning efficiency significantly in the initial stage while contributes little and even becomes harmful as the learning reward increases. This requires an adaptive integration of the optimization and learning methods during the system evolution.

\emph{4) Outer-loop ML Framework:} This work successfully and clearly show the feasibility and benefits of an integration of optimization methods and the DRL framework for efficient beamforming design. The customization of other optimization-driven ML approaches, or a combination of them, is worth of further investigation for IRS-assisted systems and also other wireless systems.

\section{Conclusion}\label{sec_con}

In this article, we have focused on the recent uses of IRS in wireless networks, and reviewed the current applications of ML approaches in IRS-assisted systems. An inspection on the common limitations of existing ML approaches motivates us to design a novel optimization-driven DRL framework for the joint beamforming optimization problem. Numerical results demonstrate that it can improve the learning efficiency and reward performance significantly compared to the conventional model-free DRL method. In the last section, we have outlined a few open research issues, which are left for our future explorations.

\bibliographystyle{IEEEtran}
\bibliography{irsmagaz}

\end{document}